%% file: main.tex
\documentclass[conference,flushend]{iaria} % Letter Size (8.5" x 11")
\pdfoutput=1 % pdflatex hint for arxiv.org (within first 5 lines)

\usepackage[babel=true,english=american]{csquotes}
\usepackage[USenglish]{babel}
\usepackage[alwaysadjust]{paralist}
\usepackage[
  babel=true, % Enable language-specific kerning
  expansion=alltext,
  protrusion=alltext-nott, % Ensure no changes at the edge of the listing.
  nopatch=eqnum, % fix unable to apply patch eqnum
  final % Always enable microtype, even if in draft mode.
]{microtype}
\usepackage{upquote}
\usepackage[zerostyle=b,scaled=.9]{newtxtt}
\DisableLigatures{encoding = T1, family = tt* }
\usepackage[shortcuts]{extdash} % Use \-/ for a breakable dash that does not prevent the remainer of the word to be hyphenated
\usepackage{relsize} % allows for \textsmaller{..}

\usepackage{fontawesome} % i.a., \faWarning{}
\usepackage{lipsum}      % for blindtext
\usepackage{booktabs}
\usepackage{amsmath,amssymb,amsfonts} % must be loaded before cleveref
\usepackage{stfloats} % floats in a twocolumn document with [t] and/or [b]
\fnbelowfloat % Put footnotes below floats
\usepackage[bb=boondox,bbscaled=.95,cal=boondoxo]{mathalpha}
\usepackage[colorinlistoftodos,prependcaption,textsize=tiny]{todonotes}

\usepackage{array}
\usepackage{enumitem}

\usepackage{xurl} % allows URLs to break on all alphanumerical chars

\vbadness=\maxdimen
\usepackage{algorithm}
\usepackage{algpseudocode}

\usepackage{makecell}

\usepackage{acro}
\acsetup{
    single = true,     % WICHTIG: Wenn ein Akronym nur 1x vorkommt -> nur Langform!
}

\input{acronyms}

\usepackage[capitalise,nameinlink,noabbrev]{cleveref} % moved to end

\usepackage{tikz}
\usetikzlibrary{
    positioning,
    arrows.meta,
    shapes.geometric,
    shapes.symbols
}

\title{A Cloud-Native Architecture for Human-in-Control LLM-Assisted OpenSearch in Investigative Settings}

\author{
  \IEEEauthorblockN{
    ~\\[-0.3ex]
    Benjamin Puhani\IEEEauthorrefmark{1},
    Kai Brehmer\IEEEauthorrefmark{1},
    Malte Prieß\IEEEauthorrefmark{2}\;\orcidlink{0009-0004-4626-2513}\\[0.3ex]~
  }
  \IEEEauthorblockA{\IEEEauthorrefmark{1}%
    AI Research Unit of the\\
    State Police of Schleswig-Holstein, Kiel, Germany\\
    e-mail: {\tt$\lbrace$benjamin.puhani\,|\,kai.brehmer$\rbrace$@polizei.landsh.de}
  }
  \\[-1.3ex]~
  \IEEEauthorblockA{\IEEEauthorrefmark{2}%
    Faculty of Computer Science and Electrical Engineering \\
    Kiel University of Applied Sciences, Germany \\
    e-mail: \tt malte.priess@haw-kiel.de}
}

\ifpdf
\pdfoutput=1 % we are running pdflatex 
\pdftrue
\fi

\makeatletter
\def\ps@IEEEtitlepagestyle{
	\def\@oddfoot{\mycopyrightnotice}
	\def\@evenfoot{}
}
\def\mycopyrightnotice{
	{\footnotesize
		\begin{minipage}{0.8\textwidth}
			\centering
			Please cite as: Puhani, B; Brehmer, K. and Prie{\ss}, M. (2026): "A Cloud-Native Architecture for Human-in-Control LLM-Assisted OpenSearch in Investigative Settings" in Proceedings of the 17th International Conference on Cloud Computing, GRIDs, and Virtualization ({CLOUD COMPUTING} 2026), p. 62-65, Lisbon, Portugal.
		\end{minipage}
    }
}

\makeatother
\begin{document}

\maketitle

\input{00_abstract}


\input{01_introduction}
\input{02_related_work}
\input{03_system_architecture}
\input{04_implementation_status}
\input{05_conclusion}

% ======== References =========
\begingroup
\sloppy
\printbibliography[notcategory=selfref]
\endgroup 

\end{document}

%% file: acronyms.tex
\DeclareAcronym{dsl}{
    short = DSL,
    long = Domain-Specific Language
}
\DeclareAcronym{soa}{
    short = SOA,
    long = Service-Oriented Architecture ,
    short-plural = s,
    long-plural = s
}
\DeclareAcronym{spa}{
    short = SPA,
    long = Single Page Application,
    short-plural = s,
    long-plural = s
}
\DeclareAcronym{ssr}{
    short = SSR,
    long = Server-Side Rendering
}
\DeclareAcronym{llm}{
    short = LLM,
    long = Large Language Model,
    short-plural = s,
    long-plural = s
}
\DeclareAcronym{ir}{
    short = IR,
    long = Information Retrieval
}
\DeclareAcronym{rag}{
    short = RAG,
    long = Retrieval-Augmented Generation
}
\DeclareAcronym{cot}{
    short = CoT,
    long = Chain-of-Thought
}
\DeclareAcronym{poc}{
    short = PoC,
    long = Proof of Concept
}

%% file: 00_abstract.tex
\begin{abstract}

Complex criminal investigations are often hindered by large volumes of unstructured evidence and by the semantic gap between natural language investigative intent and technical search logic. To address this challenge, we present a design and feasibility study of a cloud-native microservice architecture tailored to private-cloud deployments, contributing to research in secure cloud computing and leveraging modern cloud paradigms under high security and scalability requirements. The proposed system integrates \aclp{llm} into a \enquote{Human-in-Control} workflow that translates natural-language queries into syntactically valid OpenSearch \acl{dsl} expressions. We describe the implementation of a hybrid retrieval strategy within OpenSearch that combines BM25-based lexical search with nested semantic vector embeddings. The paper focuses on system design and preliminary functional validation, establishing an architectural baseline for future empirical evaluation. Technical feasibility is demonstrated through a functional prototype, and a rigorous evaluation methodology is outlined using the Enron Email Dataset as a structural proxy for restricted investigative corpora.

\end{abstract}
\acresetall % reset the acronym counter

%% file: 01_introduction.tex
\section{Introduction}
\label{sec:introduction}

Investigations into international and transnational crimes, such as genocide, crimes against humanity, and related violations of international criminal and humanitarian law, require the analysis of large volumes of unstructured textual evidence, including witness statements, interview transcripts, and communication records.
Proceedings under national universal jurisdiction frameworks, such as the German Code of Crimes against International Law (Völkerstrafgesetzbuch, VStGB), serve as a representative example of this broader class of investigations.
Across such contexts, investigators face a recurring challenge: relevant evidence is often present in the data but remains difficult to access because of the gap between natural language investigative intent and the technical logic of search systems.
Recent scholarship underscores this urgency:
\textcite{Skipanes.2025} identify the processing of unstructured text as a critical bottleneck in digital forensics and highlight that current methodologies largely fail to bridge the divide between computational opportunities and practical investigative reasoning.

Although modern search engines, such as OpenSearch~\autocite{OpenSearch}, provide scalable indexing and retrieval capabilities, they inherently require input in a rigid and structured Query \ac{dsl} rather than in natural language.
Most investigators and legal practitioners lack this expertise, leading them to rely on manual review or simple keyword searches.
These approaches are poorly suited to capturing semantic variation, indirect references, variant spellings, and translation artifacts, and they limit recall - the ability to retrieve all relevant information - precisely in those cases where exploratory and hypothesis-driven search is required.

This paper addresses this semantic gap by presenting an exploratory proof-of-concept system that integrates \acp{llm} as a translation layer between investigative intent and open search query logic.
Natural language questions are mapped to syntactically valid OpenSearch \ac{dsl} queries within a Human-in-Control architecture, in which the LLM functions as a supervised assistant rather than as an autonomous agent.
The contribution of this paper is to outline a principal system design and methodological foundation, along with a novel architectural integration that provides a basis for future empirical evaluation.
Because of legal and ethical constraints associated with real investigative data, the approach is demonstrated using the Enron Email Dataset~\autocite{Klimt.2004} as a structural proxy that exhibits key characteristics of investigative corpora, including unstructured text, noisy data, and complex communication networks.
Architecturally, the system is positioned within cloud-native computing paradigms, addressing challenges in private-cloud orchestration, horizontally scalable components, and secure cloud environments to ensure strict data sovereignty.

The remainder of this paper contextualizes this architecture within existing research (\cref{sec:related_work}), details the system design and hybrid retrieval strategy (\cref{sec:architecture}), demonstrates its functional feasibility (\cref{sec:implementation}), and outlines the roadmap for empirical evaluation (\cref{sec:conclusionAndFutureWork}).

%% file: 02_related_work.tex
\section{Related Work}
\label{sec:related_work}
%This section situates the proposed architecture within three interconnected bodies of literature.
To address the semantic and technical challenges of forensic data analysis, our work builds upon and integrates research from three primary domains.

%\textit{Bridging the Semantic Gap in Digital Forensics:}
\subsection{Bridging the Semantic Gap in Digital Forensics:}
In a recent comprehensive analysis, \textcite{Skipanes.2025} identify the processing of unstructured text as a critical bottleneck in contemporary criminal investigations.
They highlight that, while computational opportunities exist, current methods largely fail to bridge the gap between technical retrieval logic and the qualitative reasoning required by investigators.
Our work directly addresses this architectural gap by operationalizing these opportunities within a secure on-premises environment.

%\textit{LLM-Assisted Retrieval:}
\subsection{LLM-Assisted Retrieval:}
The integration of \acp{llm} into \ac{ir} systems has evolved rapidly from simple re-ranking tasks to complex query generation~\autocite{Zhu.2023}.
Current approaches often focus on \textit{Text-to-SQL} paradigms, in which \acp{llm} translate natural language into structured SQL queries for relational databases~\autocite{Shi.2024}.
However, these methods are inherently constrained by the unstructured and fuzzy nature of forensic text data.
Conversely, \ac{rag} grounds \ac{llm} responses in retrieved search results but may introduce hallucinations or lack the deterministic precision required for rigorous investigative filtering~\autocite{Gao.2024}.

%\textit{Cognitive Architectures and Prompting:}
\subsection{Cognitive Architectures and Prompting:}
Our work builds upon the findings of \textcite{Liu.2023} concerning the \enquote{Lost-in-the-Middle} phenomenon, which posits that LLMs often fail to identify relevant information in long contexts.
We address this limitation by segmenting documents into semantic units rather than processing entire texts.
Furthermore, we adopt the \ac{cot} prompting strategy proposed by \textcite{Wei.2022} to improve the logical consistency of generated OpenSearch queries.
Unlike autonomous agents, our architecture enforces a \enquote{Human-in-Control} design that prioritizes the investigator's control over search logic to ensure procedural accountability within legal domains.

%% file: 03_system_architecture.tex
\section{System Architecture and Methodology}
\label{sec:architecture}

To meet the high security and scalability requirements of law enforcement agencies, the proposed system is designed as a cloud-native microservice architecture.
While leveraging modern cloud paradigms such as containerization and orchestration, the system is intended for deployment within a restricted private-cloud environment (e.g., on-premise Kubernetes) to ensure strict data sovereignty.
At the same time, the architecture remains deployment-agnostic: the identical microservice stack can operate either in a fully orchestrated Kubernetes environment for large-scale installations or in a lightweight Docker Compose configuration for resource-constrained agencies.
This flexibility stems from consistent containerization of all services and a clear separation between application logic and infrastructure orchestration, enabling the system to scale operational complexity according to organizational needs.

\subsection{Cloud-Native Service Orchestration}
\label{subsec:orchestration}

The system follows a microservice architectural pattern composed of four distinct layers that communicate via RESTful APIs and asynchronous message queues (see~\cref{fig:architecture}): \vspace{0.1cm}

\begin{figure}[htbp]
\centering
\includegraphics[width=1\linewidth]{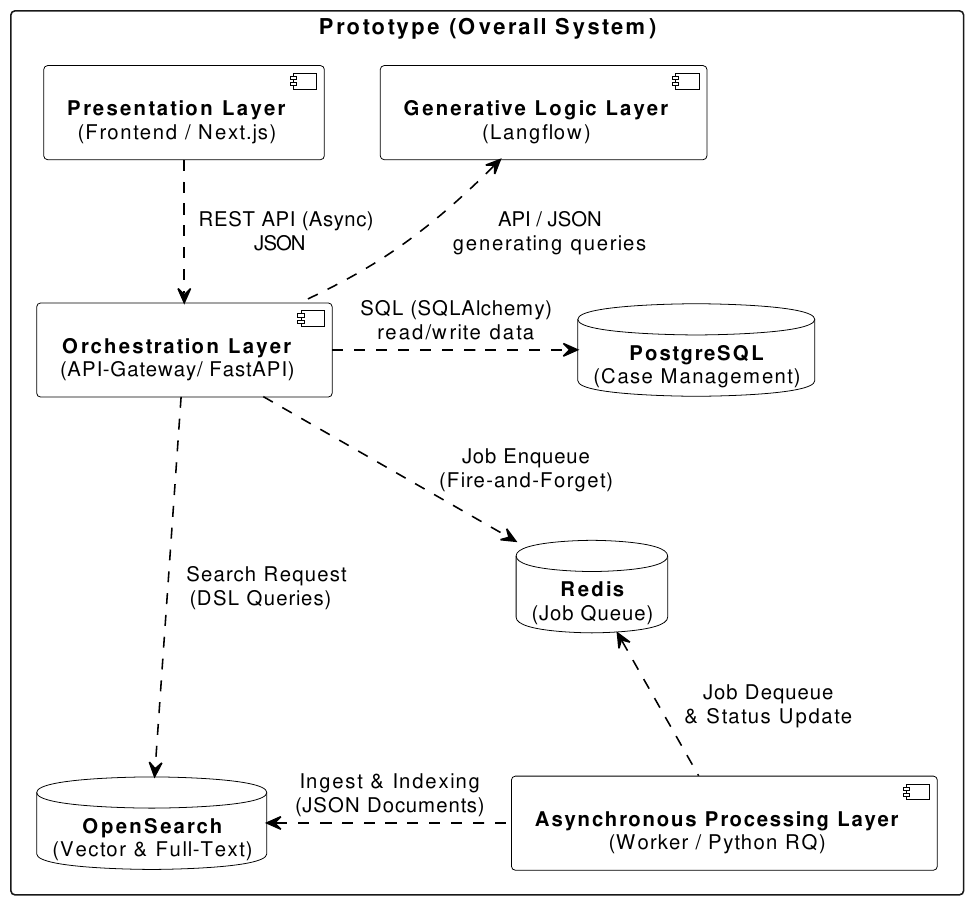}
%https://www.planttext.com?text=dLLDazeu4BrJnN_evJ210wOoIIuSjeRCrv8YWG5cSfcBOlSOBq8YaZp4jRN_FK-M3NY6v10EbDpgVg_L_Lgwi4uObszaksNNcTeA8pOKwyrMArPkuGh9XDtpu-sjYDSYvR1fE7P2fPAFFH9-4hbqI_xXHZ9BrGPO5EEFJRlLRhsXVhzFFZvJgL-V9AV_sYt2xoFeKgDpbLnggGrzNMMEgorj4ZR1_572BvX7nkXTm_f_8plXn3TX-Sy86b0YTjZvKvow4YEFOaCXtj3b2lNaX3hNPfCv9zbqo6dgZDKpMvUb0iwTTYjGDwoFgdjlsVGw92pvrL6n2-juWpoGdJ35AwwPqKwxOalTMxPYuspftmlCzIPJT_hCHaXPuVMGHwM6tthv2eL2X6mcec13RuTqOpH4ePAGZtSuj1-0RB4GBjCgXE20-uW1VO6yedzjE6vjFqLqFYnxyokUgObNgA5f4ercu_wjSBmJnO1kX7Km1AeAwnJJko5DT9h5T4PVjLd3SehmbXLxjcUkl6lE2Rhn9FKkyDHVh-97jb3n2WVMkIMKFMPh_IJLC1NrW6Q5M-58yxi06Eodydu_f1aRcw5TAkP3qljHwSmvoIomF-blT9TppW6oDB_qdMdhKiEBkma2Bhtfir3GYBy6GbZbOUjCzZfyuBMNzHazQ2aT6XmK-C1nCD_aKlRzlT9vIJhTibgmG2z1-je_y1wS6WCuPraslvf1RtuRbHD4OtzfOTQmsnYOO5zxUpsUUqVaLiP69ajNBlhx8FCX_RDSpXPeJGFN8vz71pVWxKMsDqQ9tYc0rL8UqlnwiIJ8ahgb8dgFwjDY-YN0lOlsZgN6sgtwSvzLtGGWeLNKnTz8OXGs1P0CYsImCxZ1oXeVXLTfrEf4j4zXG9xx4SK9seOSl-TiCj_Zbw67BdZgi9WptAr3xwuM4wym7zeBPtaVKNb0cZw1KIGUlBjSCMKO2OFdn-5DCewUqF92vuOSvZcpV_LUKvTw1Ijv7L-hxrx9_bR93FU5IletsgJi45ZDK9HlZw9yq1LNGMTumOJ31DvlKI4EgNw8wi4_pEK1hAvyem1ZrCmwQ7Mi4lu1Vpzbg2LTwJZt8nFAS85Zu2N-2G00
\caption{Schematic representation of the modular architecture and data flow.}
\label{fig:architecture}
\end{figure}

%\begin{enumerate}
\textit{Presentation Layer:}
A \ac{spa} built with Next.js provides the investigative user interface.
It uses \ac{ssr} to optimize initial load performance and communicates asynchronously with the backend to ensure a non-blocking user experience, which is essential for reviewing large document sets.

\textit{Orchestration Layer (API Gateway):}
The core application logic is handled by a FastAPI backend.
Unlike monolithic frameworks, FastAPI implements the ASGI standard, enabling native asynchronous request handling.
This capability is critical for maintaining high throughput when coordinating I/O-intensive operations across the database, search engine, and \ac{llm} services.
State management is offloaded to persistent stores - PostgreSQL for case management and Redis for job queues.
Crucially, this design transforms the system from a stateless search engine into a case-management workspace.
The backend tracks the \enquote{reviewed status} of each retrieved document, supporting a coverage-oriented workflow that enables investigators to systematically examine evidence by relevance.

\textit{Asynchronous Processing Layer (Worker):}
To handle large-scale data ingestion, we implemented a producer-consumer pattern using Redis.
Python-based workers perform CPU-intensive heuristic parsing and disentanglement.
However, computationally expensive vectorization is offloaded to the OpenSearch cluster thorugh a dedicated ingest pipeline that runs on specialized machine learning nodes.
This design separates the cleaning logic from the inference workload, enabling independent scaling of ingestion workers and neural inference resources.

\textit{Generative Logic Layer:}
Instead of hardcoding prompt logic, the system integrates Langflow~\autocite{Langflow} as a visual low-code environment to manage interaction chains.
To enforce data sovereignty, the architecture deliberately avoids reliance on public APIs.
Instead, it uses locally hosted open-weight models (e.g., Llama 3 or Mixtral) running on on-premise inference servers via vLLM.
This design ensures that no sensitive investigative intent leaves the secure private-cloud perimeter.
% \end{enumerate}

\subsection{Semantic Segmentation and Ingestion Strategy}
\label{subsec:segmentation}

A major challenge in processing large-scale forensic data is the \enquote{Lost-in-the-Middle} phenomenon, in which \acp{llm} fail to retrieve relevant information embedded in long, unstructured contexts~\autocite{Liu.2023}.
Indexing a typical investigative document (e.g., a 50-page witness statement or an extended email thread) as a monolithic block degrades vector search performance because of the architectural token limitations of the underlying transformer models~\autocite{Reimers.2019}.

To address this limitation, we developed a configurable, modular adapter pattern within the asynchronous worker nodes.
Unlike generic chunking strategies (e.g., fixed-size splitting), our system supports custom parsing profiles explicitly designed for each input type to preserve semantic boundaries in forensic data: \vspace{0.1cm}

% \begin{itemize}
    \textit{Heuristic Chunking (Legacy Documents):}
    For unstandardized text documents, we implemented a custom regex-based heuristic to detect semantic shifts (e.g., speaker changes or timestamps). This approach allows the system to generate atomic segments even in the absence of structured separators.
    
    \textit{Disentanglement (Communication Data):}
    For the Enron dataset, the ingestion logic was tailored to disentangle forwarded message chains.
    By selectively stripping technical headers (e.g., \texttt{X-UID}) from the semantic payload (Level 2) while preserving them for structured filtering (Level 1), we minimize noise in the vector space.
% \end{itemize}

This pre-processing step ensures that embeddings represent specific statements rather than diluted document-level averages.

\subsection{Hybrid Data Modeling and Indexing}
\label{subsec:indexing}

A core architectural decision was to implement a hybrid search index within OpenSearch.
The current approach focuses exclusively on text-based data.
Image and audio-based data would need to be converted into text.
The hybrid search index addresses the fundamental linguistic limitations of purely lexical retrieval, specifically synonymy and polysemy~\autocite{Manning.2008}.
By integrating vector-based semantic retrieval, our schema combines the strengths of both paradigms, a pattern applicable to both interview protocols and digital communication: \vspace{0.1cm}

% \begin{itemize}
    \textit{Unstructured Baseline (Level 1):}
    The document's full text and metadata are indexed using standard BM25 lexical search~\autocite{Robertson.1995}.
    This configuration ensures high recall for specific keywords, such as \textit{names, case numbers, or senders/receivers.}
    
    \textit{Structured Nested Embeddings (Level 2):}
    To mitigate context dilution, we avoid embedding long documents as a single vectors.
    Instead, the atomic semantic units identified in \cref{subsec:segmentation} (e.g., specific paragraphs, message bodies, or individual statements) are stored as nested objects containing the segment text and a 384-dimensional vector embedding.
    We use the HNSW algorithm~\autocite{Malkov.2020} with Cosine Similarity, adhering to the optimization objective of the underlying \texttt{paraphrase-multilingual-MiniLM-L12-v2} model~\autocite{Reimers.2019}.
    To minimize architectural complexity, this model is provisioned through the OpenSearch ML Commons framework and executed directly on the cluster's internal ML nodes.
    This configuration ensures that ranking is determined by semantic alignment (vector orientation) rather than by vector magnitude.
    Additionally, a search-time synonym graph expands queries, e.g., mapping \enquote{detention} to \enquote{imprisonment}, without increasing the physical index size.
% \end{itemize}

This nested structure is designed to prevent \enquote{cross-object matching} errors in which a query matches unrelated parts of a document (e.g., matching a suspect's name from page 1 with an action described on page 10), and to enable the \ac{llm} to generate queries that target specific semantic segments.

\textit{Hybrid Fusion for Prioritized Review:}
During the exploratory evidence-review phase, minimizing \textit{False Negatives} is paramount.
Investigators require a ranking mechanism that surfaces the most relevant documents first to support the coverage-oriented workflow described in \cref{subsec:orchestration}.
Our system employs a score normalization approach to fuse unbounded lexical scores (BM25) with normalized semantic scores (Cosine Similarity).
This design ensures that a document describing \enquote{off-balance sheet debt} (semantic match) is ranked competitively with documents containing the specific project code \enquote{Raptor} (lexical match), even when their vocabularies do not overlap.

\subsection{The \enquote{Human-in-Control} Generative Pipeline}
\label{subsec:pipeline}

The translation of natural language into the complex OpenSearch Query \ac{dsl} is handled by a multi-agent \ac{llm} pipeline orchestrated via Langflow.
To ensure domain agility without code modifications, the system employs schema-aware prompting: the current index definition is injected into the prompt context at runtime.
Crucially, this design enforces a strict separation of concerns:
the \ac{llm} operates exclusively on the abstract index schema, never on the actual evidentiary content.
Unlike standard \ac{rag} workflows~\autocite{Lewis.2020}, which require feeding retrieved text into the model's context window, our architecture ensures that sensitive document payloads remain confined within the OpenSearch cluster and are never exposed to the inference context.
Rather than relying on opaque \enquote{black box} logic, we implement a \ac{cot} workflow: \vspace{0.1cm}

% \begin{enumerate}
    \textit{Reasoning \& Generation:}
    The first agent acts as a \enquote{Query Architect}.
    It analyzes the user's intent and the injected schema structure, generating an intermediate reflection before constructing the JSON query.
    
    \textit{Auditing (Quality Assurance):}
    The second agent, the \enquote{Auditor}, validates each generated query against known error patterns.
    For instance, it detects when the model attempts to search for structured entities (e.g., specific dates or person names) within the semantic vector field, which typically yields lower precision.
    The Auditor enforces correct mapping to structured fields (e.g., moving a name search to the \texttt{sender} or \texttt{people} field) before execution.
    
    \textit{Transparent Execution \& Deterministic Retrieval:}
    The validated query is then executed to provide immediate feedback.
    However, unlike fully autonomous agents, the system enforces transparency:
    the generated search logic (the \enquote{translation}) is displayed alongside the results.
    Since all presented results are deterministic database retrievals rather than \ac{llm}-generated text, the risk of evidence hallucination is eliminated.
    The investigator retains full authority to evaluate the relevance of retrieved documents and iteratively refine the generated query logic, ensuring that the final assessment of evidence remains a human decision.
% \end{enumerate}

This pipeline ensures that the resulting \ac{dsl} query is syntactically valid and semantically aligned with the investigator's intent prior to execution.

%% file: 04_implementation_status.tex
\section{Implementation Status and Preliminary Feasibility}
\label{sec:implementation}

The architecture described in \cref{sec:architecture} has been implemented as a fully functional prototype.
The system successfully orchestrates interactions among the React frontend, the FastAPI gateway, and the asynchronous worker nodes.
Initial functional tests confirm that the Langflow-based \enquote{Human-in-Control} pipeline is capable of generating syntactically valid OpenSearch \ac{dsl} queries from natural language input.
Specifically, the multi-agent setup (Generator and Auditor) demonstrated the ability to detect and correct basic logical errors, such as mapping named entities to incorrect fields, before execution.
This technical readiness establishes the necessary baseline for the empirical evaluation strategy outlined in \cref{sec:conclusionAndFutureWork}, which will be the focus of subsequent research.

%% file: 05_conclusion.tex
\section{Conclusion and Future Work}
\label{sec:conclusionAndFutureWork}

This paper presented a cloud-native, microservices-based architecture designed to reduce technical barriers in accessing mass data for criminal investigations.
By combining a hybrid OpenSearch index with a supervised \ac{llm} pipeline, we established a framework to translate investigative intent into high-precision database queries without compromising data sovereignty.

%\subsection*{Planned Evaluation and Research Hypotheses}
%\label{subsec:planned-evaluation}

The next phase focuses on the quantitative calibration of the system.
As real-world investigative data is legally restricted to operational use under strict purpose limitation regulations, it cannot be utilized for public academic benchmarking.
Therefore, we will utilize the Enron email dataset~\autocite{Klimt.2004} as a reproducible Ground Truth to simulate forensic retrieval tasks.

The primary objective is not merely to compare algorithms, but to determine the optimal hybrid configuration for forensic workflows, which prioritize high recall (coverage) over precision.
The evaluation will specifically investigate two core architectural decisions: \vspace{0.1cm}

% \begin{enumerate}
    \textit{Segmentation Granularity:}
    Comparing retrieval performance when indexing monolithic documents versus the proposed granular segmentation (e.g., heuristic chunking or thread-splitting).
    We hypothesize that granular segments yield higher relevance scores for specific details but require aggregation to preserve context.
    
    \textit{Score Fusion Tuning:}
    Evaluating different weighting strategies for the Score Normalization (Lexical vs. Semantic weights) to maximize Recall@100.
    This metric, which measures the proportion of relevant documents retrieved within the top 100 results~\autocite{Manning.2008}, is chosen to reflect the operational reality of investigators, who require the most critical evidence to appear within the first few pages of results.
% \end{enumerate}

\noindent To test these hypotheses, the evaluation design involves: \vspace{0.1cm}

% \begin{itemize}
    \textit{Adversarial Scenario Design:}
    We will define 5--10 distinct search scenarios designed to simulate the semantic gap.
    Instead of searching for known identifiers (e.g., \enquote{Project Raptor}), queries will formulate abstract investigative intents (e.g., \enquote{conversations expressing anxiety about the company's financial stability} or \enquote{instructions to destroy documents}).
    These scenarios are specifically chosen to challenge lexical search engines, as they rely on sentiment and context rather than unique keywords.

    \textit{Semantic Ground Truth Construction:}
    We utilize the publicly available CMU Enron Corpus~\autocite{Klimt.2004} as the structural baseline.
    For the evaluation of retrieval performance (Recall/Precision), we utilize established relevance judgments from the TREC Legal Track or comparable academic annotation sets (e.g., UC Berkeley Enron Analysis).
    This ensures that the Ground Truth represents the semantic reality of the documents, independent of the specific vocabulary used in the query.

    \textit{Comparative Ablation Study:}
    To quantify the added value of the hybrid architecture, we will conduct an ablation study comparing three configurations:
    \begin{enumerate}[label=(\alph*)]
        \item Purely Lexical (Level 1 BM25),
        \item Purely Semantic (Level 2 Vector-only), and
        \item Hybrid Score Fusion (Level 1 + Level 2).
    \end{enumerate}
% \end{itemize}

We hypothesize that the hybrid system yields the highest Recall@100, demonstrating superior robustness in scenarios where suspects employ obfuscated language or indirect phrasing that evades purely lexical detection.

% Future work will also address the integration of the prototype into the secure, air-gapped infrastructure of the State Police Office to validate user acceptance among domain experts.

% \subsection{Addressing Domain Shift via Synthetic Data}
% \label{subsec:future-synthetic}

% While the Enron dataset serves as a robust structural proxy for unstructured mass data, we acknowledge the semantic gap between corporate communication and international criminal investigations.
% Forensic evidence in the target domain often features distinct linguistic characteristics, such as descriptions of violence, multilingual code-switching, and vague temporal references in witness statements.
% To validate the system's semantic capabilities without compromising classified operational data, future work will implement a privacy-preserving evaluation, using synthetic datasets.
% We plan to utilize \acp{llm} to generate synthetic witness statements and interview protocols that replicate the specific ontology of international criminal law (e.g., adhering to definitions in the Rome Statute), but populated with entirely fictitious entities.
% This will allow us to benchmark the intent translation logic against complex legal concepts (e.g., \enquote{Command Responsibility}) prior to the final deployment on air-gapped infrastructure.

%% file: literature.bib
@book{Manning.2008,
  author    = {Manning, Christopher D. and Raghavan, Prabhakar and Schütze, Hinrich},
  title     = {Introduction to Information Retrieval},
  publisher = {Cambridge University Press},
  address   = {Cambridge},
  year      = {2008},
  isbn      = {9780521865715},
  doi       = {10.1017/CBO9780511809071}
}

@article{Liu.2023,
    title = "Lost in the Middle: How Language Models Use Long Contexts",
    author = "Liu, Nelson F.  and
      Lin, Kevin  and
      Hewitt, John  and
      Paranjape, Ashwin  and
      Bevilacqua, Michele  and
      Petroni, Fabio  and
      Liang, Percy",
    journal = "Transactions of the Association for Computational Linguistics",
    volume = "12",
    year = "2024",
    address = "Cambridge, MA",
    publisher = "MIT Press",
    doi = "10.1162/tacl_a_00638",
    pages = "157--173",
    abstract = "While recent language models have the ability to take long contexts as input, relatively little is known about how well they use longer context. We analyze the performance of language models on two tasks that require identifying relevant information in their input contexts: multi-document question answering and key-value retrieval. We find that performance can degrade significantly when changing the position of relevant information, indicating that current language models do not robustly make use of information in long input contexts. In particular, we observe that performance is often highest when relevant information occurs at the beginning or end of the input context, and significantly degrades when models must access relevant information in the middle of long contexts, even for explicitly long-context models. Our analysis provides a better understanding of how language models use their input context and provides new evaluation protocols for future long-context language models.",
    %url = "https://aclanthology.org/2024.tacl-1.9/",
}

@article{Malkov.2020,
  author={Malkov, Yu A. and Yashunin, D. A.},
  journal={IEEE Transactions on Pattern Analysis and Machine Intelligence}, 
  title={Efficient and Robust Approximate Nearest Neighbor Search Using Hierarchical Navigable Small World Graphs}, 
  year={2020},
  volume={42},
  number={4},
  pages={824-836},
  keywords={Routing;Complexity theory;Search problems;Data models;Approximation algorithms;Biological system modeling;Brain modeling;Graph and tree search strategies;artificial intelligence;information search and retrieval;information storage and retrieval;information technology and systems;search process;graphs and networks;data structures;nearest neighbor search;big data;approximate search;similarity search},
  doi={10.1109/TPAMI.2018.2889473},
  %url = {https://arxiv.org/abs/1603.09320}, % diese Url zeigt auf eine lesbare version
}

@inproceedings{Wei.2022,
    author = {Wei, Jason and Wang, Xuezhi and Schuurmans, Dale and Bosma, Maarten and Ichter, Brian and Xia, Fei and Chi, Ed and Le, Quoc V. and Zhou, Denny},
    booktitle = {Advances in Neural Information Processing Systems},
    editor = {Koyejo, S. and Mohamed, S. and Agarwal, A. and Belgrave, D. and Cho, K. and Oh, A.},
    pages = {24824--24837},
    publisher = {Curran Associates, Inc.},
    title = {Chain-of-Thought Prompting Elicits Reasoning in Large Language Models},
    volume = {35},
    year = {2022},
    %note = {\url{https://proceedings.neurips.cc/paper_files/paper/2022/file/9d5609613524ecf4f15af0f7b31abca4-Paper-Conference.pdf}}
    %url = {https://proceedings.neurips.cc/paper_files/paper/2022/file/9d5609613524ecf4f15af0f7b31abca4-Paper-Conference.pdf},
}

@article{Zhu.2023,
 author = {Zhu, Yutao and Yuan, Huaying and Wang, Shuting and Liu, Jiongnan and Liu, Wenhan and Deng, Chenlong and Chen, Haonan and Liu, Zheng and Dou, Zhicheng and Wen, Ji-Rong},
 year = {2026},
 title = {Large Language Models for Information Retrieval: A Survey},
 pages = {1--54},
 pagination = {page},
 volume = {44},
 issn = {1046-8188},
 journaltitle = {ACM Transactions on Information Systems},
 shortjournal = {ACM Trans. Inf. Syst.},
 doi = {10.1145/3748304},
 number = {1},
 abstract = {As a primary means of information acquisition, information retrieval (IR) systems, such as search engines, have integrated themselves into our daily lives. These systems also serve as components of dialogue, question-answering, and recommender systems. The trajectory of IR has evolved dynamically from its origins in term-based methods to its integration with advanced neural models. While the neural models excel at capturing complex contextual signals and semantic nuances, thereby reshaping the IR landscape, they still face challenges such as data scarcity, interpretability, and the generation of contextually plausible yet potentially inaccurate responses. This evolution requires a combination of both traditional methods (such as term-based sparse retrieval methods with rapid response) and modern neural architectures (such as language models with powerful language understanding capacity). Meanwhile, the emergence of large language models (LLMs), typified by ChatGPT and GPT-4, has revolutionized natural language processing due to their remarkable language understanding, generation, generalization, and reasoning abilities. Consequently, recent research has sought to leverage LLMs to improve IR systems. Given the rapid evolution of this research trajectory, it is necessary to consolidate existing methodologies and provide nuanced insights through a comprehensive overview. In this survey, we delve into the confluence of LLMs and IR systems, including crucial aspects such as query rewriters, retrievers, rerankers, and readers. Additionally, we explore promising directions, such as search agents, within this expanding field.},
 file = {Zhu, Yuan et al. 2023 - Large Language Models for Information:Attachments/Zhu, Yuan et al. 2023 - Large Language Models for Information.pdf:application/pdf},
 %url = {https://arxiv.org/pdf/2308.07107},
}

@article{Shi.2024,
 author = {Shi, Liang and Tang, Zhengju and Zhang, Nan and Zhang, Xiaotong and Yang, Zhi},
 year = {2026},
 title = {A Survey on Employing Large Language Models for Text-to-SQL Tasks},
 keywords = {Computation and Language (cs.CL)},
 pages = {1--37},
 pagination = {page},
 volume = {58},
 issn = {0360-0300},
 journaltitle = {ACM Computing Surveys},
 shortjournal = {ACM Comput. Surv.},
 doi = {10.1145/3737873},
 number = {2},
 abstract = {With the development of the Large Language Models (LLMs), a large range of LLM-based Text-to-SQL(Text2SQL) methods have emerged. This survey provides a comprehensive review of LLM-based Text2SQL studies. We first enumerate classic benchmarks and evaluation metrics. For the two mainstream methods, prompt engineering and finetuning, we introduce a comprehensive taxonomy and offer practical insights into each subcategory. We present an overall analysis of the above methods and various models evaluated on well-known datasets and extract some characteristics. Finally, we discuss the challenges and future directions in this field.},
 file = {A Survey on Employing Large Language Models for Text-to-:Attachments/A Survey on Employing Large Language Models for Text-to-.pdf:application/pdf},
 %url = {https://arxiv.org/pdf/2407.15186},
}

@inproceedings{Reimers.2019,
    title = "Sentence-{BERT}: Sentence Embeddings using {S}iamese {BERT}-Networks",
    author = "Reimers, Nils and Gurevych, Iryna",
    booktitle = "Proceedings of the 2019 Conference on Empirical Methods in Natural Language Processing",
    year = "2019",
    publisher = "Association for Computational Linguistics",
    doi = {10.48550/arXiv.1908.10084},
    %month = "11",
    %url = "http://arxiv.org/abs/1908.10084",
}

@inproceedings{Robertson.1995,
  title = {Okapi at TREC-3},
  author = {Robertson, Stephen E and Walker, Steve and Jones, Susan and Hancock-Beaulieu, Micheline M and Gatford, Mike},
  booktitle = {Overview of the Third Text REtrieval Conference (TREC-3)},
  pages = {109--126},
  year = {1995},
  publisher = {NIST},
  % note = {\url{https://trec.nist.gov/pubs/trec3/t3_proceedings.html}},
  %url = {https://trec.nist.gov/pubs/trec3/t3_proceedings.html},
  %urldate = {2026-01-25}
}

@InProceedings{Klimt.2004,
    author="Klimt, Bryan
    and Yang, Yiming",
    editor="Boulicaut, Jean-Fran{\c{c}}ois
    and Esposito, Floriana
    and Giannotti, Fosca
    and Pedreschi, Dino",
    title="The Enron Corpus: A New Dataset for Email Classification Research",
    booktitle="Machine Learning: ECML 2004",
    year="2004",
    publisher="Springer Berlin Heidelberg",
    address="Berlin, Heidelberg",
    pages="217--226",
    isbn="978-3-540-30115-8",
    doi= {10.1007/978-3-540-30115-8_22}
}

@software{OpenSearch,
  author       = {{OpenSearch Project}},
  title        = {{OpenSearch}},
  version      = {3.4},
  year         = {2025}, % Das jahr wo die Version 3.4 released wurde
  organization = {The Linux Foundation},
  % Wir kommentieren das normale URL-Feld aus, damit LaTeX nicht sein eigenes Format erzwingt:
  %url = {https://opensearch.org/}, 
  %urldate = {2026-01-25},
  % Stattdessen packen wir alles exakt so, wie der Editor es will, in die Note:
  note         = {\url{https://opensearch.org/} [visited: 2026-03-13]}
}

@software{Langflow,
  author  = {{Langflow AI}},
  title   = {{Langflow}},
  version = {1.6.8},
  year    = {2025}, % Das jahr wo die Version 1.6.8 released wurde
  % url     = {https://github.com/langflow-ai/langflow},
  note    = {\url{https://github.com/langflow-ai/langflow} [visited: 2026-03-13]}
}

@article{Skipanes.2025,
    author = {Skipanes, Mads and Demartini, Gianluca and Franke, Katrin and Nissen, Alf Bernt},
    title = {Information analysis in criminal investigations: methods, challenges, and computational opportunities processing unstructured text},
    journal = {Policing: A Journal of Policy and Practice},
    volume = {19},
    pages = {paaf005},
    year = {2025},
    month = {03},
    issn = {1752-4520},
    doi = {10.1093/police/paaf005}
}

@misc{Gao.2024,
      title={Retrieval-Augmented Generation for Large Language Models: A Survey}, 
      author={Yunfan Gao and Yun Xiong and Xinyu Gao and Kangxiang Jia and Jinliu Pan and Yuxi Bi and Yi Dai and Jiawei Sun and Meng Wang and Haofen Wang},
      year={2024},
      doi = {10.48550/arXiv.2312.10997}
      %note= {\url{https://arxiv.org/abs/2312.10997}},
      %eprint={2312.10997},
      %archivePrefix={arXiv},
      %primaryClass={cs.CL},
      %url={https://arxiv.org/abs/2312.10997},
}

@inproceedings{Lewis.2020,
  author    = {Lewis, Patrick and Perez, Ethan and Piktus, Aleksandra and Petroni, Fabio and Karpukhin, Vladimir and Goyal, Naman and Küttler, Heinrich and Lewis, Mike and Yih, Wen-tau and Rocktäschel, Tim and Riedel, Sebastian and Kiela, Douwe},
  title     = {{Retrieval-Augmented Generation} for Knowledge-Intensive {NLP} Tasks},
  booktitle = {Advances in Neural Information Processing Systems},
  volume    = {33},
  editor    = {Larochelle, H. and Ranzato, M. and Hadsell, R. and Balcan, M.F. and Lin, H.},
  publisher = {Curran Associates, Inc.},
  pages     = {9459--9474},
  year      = {2020},
  %note      = {\url{https://proceedings.neurips.cc/paper/2020/file/6b493230205f780e1bc26945df7481e5-Paper.pdf}}
}
